\documentclass[]{article}
\usepackage{graphicx}
\usepackage{hyperref}
\usepackage{ulem}
\usepackage[section]{placeins}
\usepackage{amsmath}
\usepackage{amssymb}
\usepackage{dsfont}
\usepackage[comma,super,sort&compress]{natbib}

\newcommand{\ket}[1]{|#1\rangle}             

\makeatletter
\renewcommand\@biblabel[1]{#1.}
\makeatother

\newcommand\blfootnote[1]{%
  \begingroup
  \renewcommand\thefootnote{}\footnote{#1}%
  \addtocounter{footnote}{-1}%
  \endgroup
}

\begin{document}

\title{Full characterisation of polarisation states of light via direct measurement}
\author{Jeff~Z.~Salvail,$^{1\star}$ Megan~Agnew,$^{1}$ Allan~S.~Johnson,$^{1}$ \\ Eliot~Bolduc,$^{1}$ Jonathan~Leach,$^{1}$ and Robert~W.~Boyd$^{1, 2}$}

\maketitle

\blfootnote{$^1$Dept.~of Physics, University of Ottawa, Ottawa, Canada, $^2$Institute of Optics, University of Rochester, Rochester, USA. $^{\star}$e-mail: jeff.salvail@gmail.com}

{\bf Ascertaining the physical state of a system is vital in order to understand and predict its behaviour. However, due to their fragile nature, the direct observation of quantum states has been elusive until recently.  Historically, determination of the quantum state has been performed indirectly, through use of tomography. We report on two experiments showing that an alternative approach can be used to determine the polarisation quantum state in a simple, fast, and general manner. The first experiment entails the direct measurement of the probability amplitudes describing pure polarisation states of light, the first such measurement on a two-level system. The second experiment entails the direct measurement of the Dirac distribution (a phase-space quasi-probability distribution informationally equivalent to the density matrix), demonstrating that the direct measurement procedure is applicable to general (i.e., potentially mixed) quantum states. Our work has applications to measurements in foundational quantum mechanics, quantum information, and quantum metrology.

}

Measurement plays a vital role in the practice of science. 
This is especially so in the case of quantum mechanics, where the measurement process is fundamental to the formulation of the theory.
A crucial feature of quantum mechanics is that a measurement of one variable of a system erases information about the corresponding conjugate variable.
The classic example is that determining the position of a particle disturbs its momentum, and vice versa.
These measurements, known as strong measurements, collapse the wavefunction such that no additional information can be obtained.

In order to completely determine a quantum state, which is described in general by complex numbers, one must perform multiple measurements on many identical copies of the system.
Quantum tomography \cite{DAriano:2003tw} is one method of quantum state determination that uses strong measurements \cite{Banaszek:1999ur,White:2001vi,Itatani:2004wp,Resch:2005vj,Agnew:2011wj}. Tomographic reconstruction entails estimating the complex numbers that describe the state from the real-valued probabilities that result from strong measurements. Consequently, this approach can be considered \textit{indirect} state determination due to the requirement of post-processing.

The first demonstration of \textit{direct} quantum wavefunction measurement was recently reported \cite{Lundeen:2011hj}.
In this study, the transverse spatial wavefunction, that is, the probability amplitude for photon detection at each position $\psi(x)$, was measured directly.
In contrast to tomography, this method is considered direct because the measurement apparatus records the complex probability amplitudes describing the state, and therefore there is no need for post-processing.
The technique for direct quantum state determination is applicable to many different systems, which, as the authors of   \cite{Lundeen:2011hj} point out, includes the polarisation degree of freedom.
Recently it has been proposed that this technique can be generalized to measure all aspects of a general quantum state, i.e.,  so that it is compatible with mixed states \cite{Lundeen:2012db}.

Although familiar and convenient, the density matrix is not the only way to describe a general quantum state. A state can be expressed in terms of its Dirac quasi-probability distribution (or phase-space representative), which is informationally equivalent to the density matrix $\rho$ \cite{Lundeen:2012db,Dirac:1945wx, Chaturvedi:2006kz,Hofmann:2012ey}.
Quasi-probability distributions have been both studied theoretically, in the context of discrete systems \cite{Feynman:1987ur, Leonhardt:1995wb}; and measured directly, for case of the spatial Wigner function \cite{Mukamel:2003uv, Smith:2005vj}.
The Dirac distribution is particularly useful due to its relation to the direct measurement technique \cite{Lundeen:2012db}.

Directly measuring a quantum system relies on the technique of weak measurement: extracting so little information from a single measurement that the state does not collapse \cite{Aharonov:1988wha,Duck:1989vo,Knight:1990vba,Ritchie:1991vf,Wiseman:2002jt,Solli:2004ix,Resch:2004es,Johansen:2004cc,Pryde:2005gk,Hosten:2008ih,Dixon:2009cz,Popescu:2009cz,Kocsis:2011jg,Feizpour:2011bs,deGosson:2012kt}.
The first measurement of a weak value was the amplified transverse displacement between the polarisation components of light induced by a birefringent crystal \cite{Ritchie:1991vf}.  More recently, the technique was used to observe the transverse displacement of a beam of light by only several angstroms \cite{Hosten:2008ih} and an angular rotation on the order of femtoradians \cite{Dixon:2009cz}. Weak measurement was recently proposed as a tool to study nonlinear optical phenomena with single photons by amplifying the apparent photon number \cite{Feizpour:2011bs}. Weak measurements have also allowed observation of apparent super-luminal velocity \cite{Solli:2004ix} and the mapping of average photon trajectories after they pass through a double slit \cite{Kocsis:2011jg}.

The main results of our paper are the direct measurements of the wavefunction and Dirac distributions for polarisation states of light.  These results are the first direct measurements that are applicable to qubits - the fundamental unit of quantum information.
We demonstrate direct state measurement in a two-dimensional Hilbert space by weakly coupling the polarisation state of light to the spatial degree of freedom. This study extends previous work on polarisation weak measurements \cite{Duck:1989vo,Ritchie:1991vf, Pryde:2005gk}. We obtain the weak value by introducing a small spatial shift between the horizontal and vertical polarisation components, then strongly measuring the polarisation in the diagonal/anti-diagonal basis.
Importantly, our experimental implementation determines the general description of the state, and, in contrast with previous experimental work, it is not limited to pure states.

In our experimental procedure, we use direct measurement to determine the polarization state of the photons in an intense beam of light that has been prepared such that each photon is in the same quantum polarisation state.  Thus, even though the light beam is intense, our procedure determines the quantum polarization state of each photon.  We note that certain photon states, such as those that involve entanglement, could not be measured using the exact procedures described here.  For such states, the measurement would need to be performed at the single photon or biphoton level.  The basic procedure outlined in this paper could still be used in this situation, although the detection process would need to be performed using single-photon detectors.  In this regard, we note that recent work has shown that cooled \cite{Kocsis:2011jg} or commercial electron-multiplying \cite{Edgar:2012bl} CCD  cameras can be operated at the single-photon level and with sufficient sensitivity to determine quantum features of the light field.

\begin{figure*}[htbp]
\begin{center}
\includegraphics[width=\textwidth]{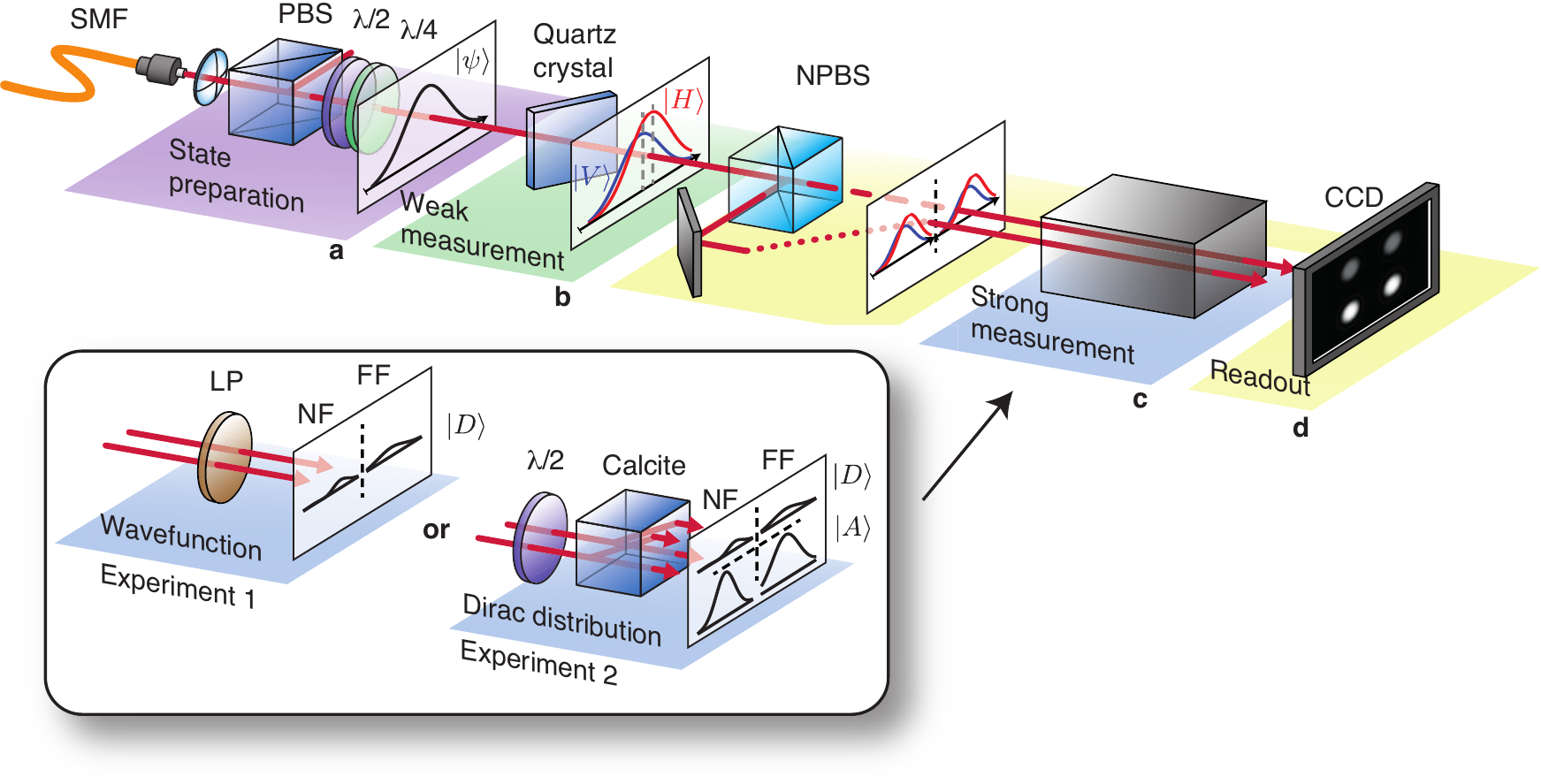}
\caption{\textbf{Schematic representation of the experiment.}
\textbf{a,} The output of the single-mode fibre (SMF) is a near-Gaussian transverse mode of light. A polarising beamsplitter (PBS) and waveplate(s) create a known pure polarisation state.
\textbf{b,} A quartz crystal at an oblique angle performs the weak measurement by introducing a small (compared to the beam waist) lateral displacement between horizontal and vertical polarisation components.
\textbf{c,} A strong measurement in a basis (diagonal/anti-diagonal) mutually unbiased from the weak measurement is used to complete the direct measurement.
\textbf{Inset:} To measure the wavefunction, a linear polariser oriented to transmit diagonally polarised light performs the strong measurement and post-selection; to measure the Dirac distribution, a $\lambda/2$ wave-plate and calcite beam displacer perform the strong measurement.
\textbf{d,} A 50:50 non-polarising beam splitter (NPBS) splits the light into two sub-ensembles. These are imaged in the near-field (dotted line, NF) and far-field (dashed line, FF) of the quartz crystal onto non-overlapping regions of the CCD camera. 
}
\label{fig:experiment}
\end{center}
\end{figure*}

\vspace{0.3cm}{\raggedleft {\it Theory.$-$} } In any quantum measurement, the observer couples an unknown probe state to a pointer that reads out the value. For example, a birefringent crystal can couple the polarisation state of light to the spatial degree of freedom;  in this case, the initial polarisation state is called the probe state, and the spatial degree of freedom of the light is considered the pointer. If the pointer state is a Gaussian mode with width $w$, a strong measurement separates the eigenstates of the measurement operator $\hat{A}$, by an amount $\delta \gg w$, such that the eigenstates are completely resolved.

Weak measurements occur in the opposite regime, where the coupling is much less than the pointer width $\delta \ll w$. In this case, the eigenstates of $\hat{A}$ are not resolved by the pointer, so the wavefunction does not collapse.  Therefore, a subsequent measurement performed on the quantum state can be used to extract further information. If the subsequent measurement is strong, such that the eigenstates are resolved, we can choose to consider only the statistics of one particular outcome; this is called post-selection and the chosen outcome of interest is the post-selected state. The average result of the weak measurement is called the weak value and is given by
\begin{align} \label{weakvalue}
\langle \hat{A} \rangle^W_{\phi} = \frac{\langle \phi | \hat{A}  \rho  | \phi \rangle}{\langle \phi |  \rho  | \phi \rangle},
\end{align}
where $\rho$ is the density operator that describes the initial state and $| \phi \rangle$ is the final, post-selected, state \cite{Wiseman:2002jt,Johansen:2004cc}. In the case that the initial state is pure and may be described by the state vector $|\psi\rangle$ (i.e., $\rho = |\psi \rangle \langle \psi |$), the weak value in equation (\ref{weakvalue}) simplifies to the form first introduced by Aharonov, Albert, and Vaidman \cite{Aharonov:1988wha}:
\begin{equation}
\langle \hat{A} \rangle^W_{\phi} = \frac{\langle \phi | \hat{A} | \psi \rangle}{\langle \phi | \psi \rangle}.
\end{equation}

In the case that $| \psi \rangle = | \phi \rangle$, the expectation value of the weak measurement is equal to the standard expectation value of the operator $\hat{A}$. In general, the initial and final states may differ, and the weak value can be complex. For the specific case where the initial and final states are nearly orthogonal, the weak value can become arbitrarily large, leading to the amplification effect discussed above. The complex nature of the weak value, combined with the fact that weak measurement does not significantly disturb the system, enables the direct measurement of the quantum state via weak measurements.

The complex weak value is determined by characterising the pointer. The pointer's position indicates the real part of the weak value $\mathrm{Re}[\langle \hat{A} \rangle^{W}_{\phi}]$, and the pointer's momentum indicates the imaginary part $\mathrm{Im}[\langle \hat{A} \rangle^{W}_{\phi}]$ \cite{Lundeen:2005jr}.

In the specific case that the weak and final measurements are of mutually unbiased\cite{{Wootters:1989uo}} variables (see Supplementary Note 1), the weak values have a direct relationship to the state description.  The coefficients $c_i$ of a wavevector $|\psi\rangle$ that describes a pure quantum state can be written in terms of specific weak values:
\begin{align} \label{wavefunction}
c_{i} &=  \langle a_i | \psi\rangle  = \nu \langle \pi_{a_i} \rangle^W_{b_j}.
\end{align}
Here the weakly measured observable $\pi_{a_i} = |a_i \rangle \langle a_i|$ is the projector into the $i^{th}$ state of the basis $\mathcal{A}$ \cite{Lundeen:2011hj}.  The factor $\nu$ is a constant of normalisation independent of $i$ and may be taken to be real. Equation (\ref{wavefunction}) shows that the wavefunction describing a pure state can be directly measured by scanning weak measurements in basis $\mathcal{A}$ and post-selecting on a fixed state in the mutually unbiased basis $\mathcal{B}$, then normalising the wavefunction.

The procedure that uses equation (\ref{wavefunction}) can be extended to give a technique to directly measure the most general description of the quantum state. 
The simplest such generalisation entails measuring weakly in basis $\mathcal{A}$, followed by recording the results of all outcomes of the strong measurement in basis $\mathcal{B}$. In terms of the density operator $\rho$, the elements of the Dirac distribution \cite{Dirac:1945wx}, which describes a general quantum state, can be written in terms of specific weak values:
\begin{align} \label{Diracdistribution}
S_{ij} = & \langle b_j | a_i \rangle  \langle a_i |  \rho | b_j \rangle =  p_{b_j} \langle \pi_{a_i} \rangle_{b_j}^W.
\end{align}
That is to say, the $(i, j)^{th}$ element of the Dirac distribution is equal to the result of the weak measurement of $\pi_{a_i}$ followed by post-selection on state $b_j$, multiplied by the probability of successful post-selection $p_{b_j}  = \langle b_j |  \rho  | b_j \rangle$ \cite{Lundeen:2012db}.  Importantly, one can always invert equation (\ref{Diracdistribution}) and calculate the density matrix $\rho$ from the measured Dirac distribution $S$.  For further details on equations~(\ref{wavefunction}) and (\ref{Diracdistribution}), see Supplementary Notes 2 and 3. 

The Dirac distribution is an underused but elegant way to describe a general quantum state. In particular, it is very useful for visualising discrete systems. In our work, we use Chaturvedi \textit{et al.}'s \cite{Chaturvedi:2006kz} ``left" phase-space representative throughout, and discuss only the discrete (i.e., $N$-level) Hilbert space version. The connection between Dirac distribution, joint probabilities and the weak value was also explored by Hofmann\cite{Hofmann:2012ey}.

An important result is that a single weak value completely determines the wavefunction of a qubit (see Supplementary Note 2).
For a single photon, the weak measurement has very large uncertainty; thus, the above procedure must be repeated on many photons, or equivalently on a classical light beam, to establish the weak value with a high degree of confidence.

\vspace{0.3cm}{\raggedleft {\it Experiment.$-$} } We perform two experiments. First, we implement the technique encapsulated by equation (\ref{wavefunction}) to measure a variety of pure polarisation wavefunctions. Second, we apply the technique encapsulated by equation (\ref{Diracdistribution}) to measure the Dirac distribution of a variety of states. The only difference between the two experiments is in the nature of the strong measurement: in the first experiment, a single strong measurement outcome is required; whereas in the second experiment, all eigenstates of the strong measurement are recorded.

A brief summary of the experimental procedures is as follows; see Fig.~\ref{fig:experiment} for a schematic. 
First, the probe (polarisation) and pointer (spatial mode) states are prepared (a). 
Second, the weak measurement is performed by a quartz plate, which slightly displaces the two orthogonal polarisation components $|H\rangle$ and $|V\rangle$ of the probe laterally (b). 
Third, the strong measurement in the $D/A$ basis is performed (c). To measure the wavefunction, we post-select the final state by projecting the polarisation into the diagonal state $|D\rangle$ using a linear polariser (LP) oriented to transmit diagonally polarised light. To measure the Dirac distribution, a calcite crystal separates the components $|D\rangle$ and $|A\rangle$ so that they do not overlap.
Finally, the wavefunction or Dirac distribution is read out by imaging the near- and far-fields of the plane immediately after the quartz onto separate regions of interest of a CCD camera (d). Two regions are used to read out the wavefunction and four are needed to read out the Dirac distribution.

In order to demonstrate our ability to perform the direct measurement of the polarisation state, we measure the probability amplitudes of three sets of input polarisation states, each corresponding to a different great circle on the Poincar\'{e} sphere. The states are created by appropriate orientation of a half-wave plate and, optionally, a quarter-wave plate.  In the second experiment, we create a number of states in the same fashion and measure their Dirac distributions, then calculate the associated density matrix.

\begin{figure}
\begin{center}
\includegraphics[width = 3in]{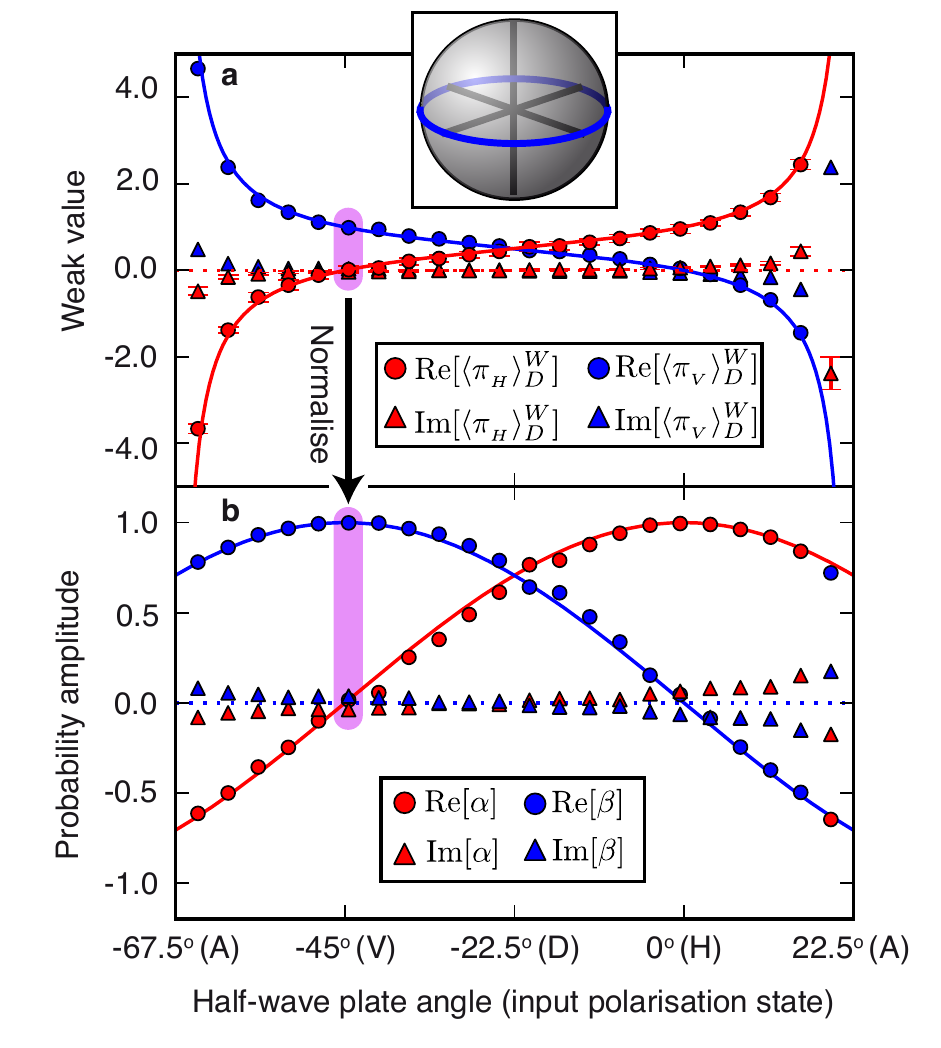}
\end{center}
\caption[weakvals]{\textbf{Results of experiment 1 with linearly polarised probe states.} \textbf{a,} Measured average weak values plotted as a function of input polarisation angle, $0^{\circ}$ is defined as parallel to the optical table. Error bars are shown only for the red points for clarity, and indicate the standard deviation of 100 independently measured weak values. \textbf{b,} Real and imaginary components of the probability amplitudes determined by normalising the weak values of each test state, where $|\psi\rangle = \alpha |H \rangle + \beta |V\rangle$. For both panels, the solid lines are the theoretical predictions of the real components, and the dotted lines are the theoretical predictions of the imaginary components. \textbf{Inset:} A Poincar\'{e} sphere with the path taken indicated by the blue line. }
\label{fig:weakvals}
\end{figure}

\begin{figure}
\begin{center}
\includegraphics[width = 3in]{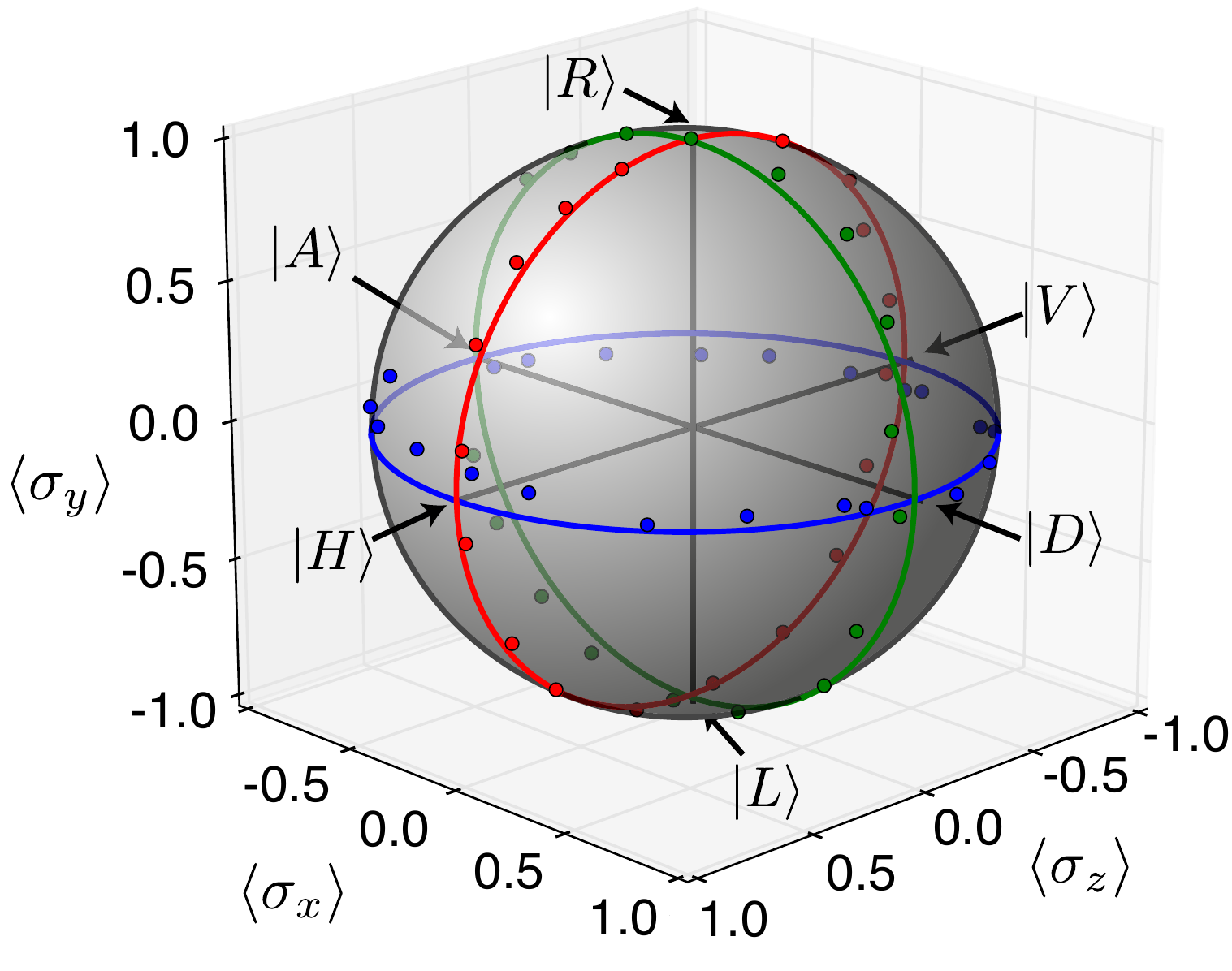}
\end{center}
\caption[weakvals]{ Measured states on the Poincar\'{e} sphere.
Here we show a Poincar\'{e} sphere with the set of directly measured states, indicated by their calculated Stokes parameters $\langle \sigma_x \rangle = \langle \psi | (\pi_D - \pi_A) | \psi \rangle$, $\langle \sigma_y \rangle = \langle \psi | (\pi_R - \pi_L) | \psi \rangle$, $\langle \sigma_z \rangle = \langle \psi | (\pi_H - \pi_V) | \psi \rangle$.
The blue points indicate states created by rotating the half-wave plate. The red (green) points indicate calculated Stokes parameters for states created by rotating the half-wave plate, followed by a quarter-wave plate at fixed angle $0^{\circ}$ ($45^{\circ}$). The solid lines indicate the paths taken for each data set.}
\label{fig:sphere}
\end{figure}

\vspace{0.3cm}{\raggedleft {\it Results.$-$} } Figure~\ref{fig:weakvals} shows the measured weak values and corresponding polarisation probability amplitudes as a function of input polarisation angle. Figure~\ref{fig:sphere} shows the calculated Stokes parameters for each measured $|\psi\rangle$ in Figure \ref{fig:weakvals} (blue points). We also show calculated Stokes parameters for two additional paths around the Poincar\'{e} sphere. Measured weak values and probability amplitudes are included in Supplementary Figures S1 and S2 for all these states.

\begin{figure}
\begin{center}
\includegraphics[width = 3in]{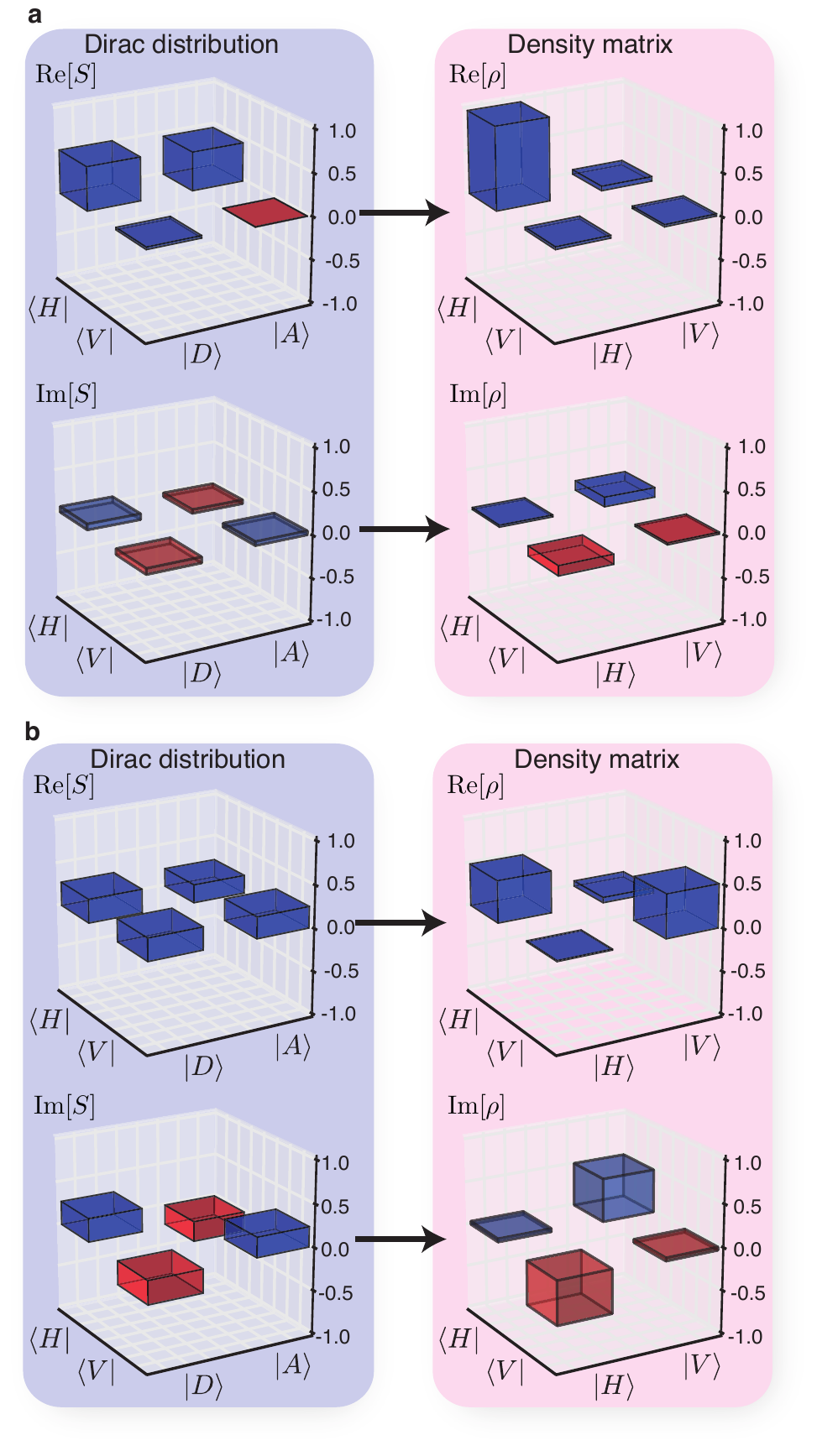}
\end{center}
\caption{\textbf{Results of experiment 2.}  The directly measured Dirac distributions and corresponding density matrices for the horizontal linear polarisation state $\ket{H}$ \textbf{a} and left-hand circular polarisation state $\ket{L} = 1/{2}((1+i)\ket{H} + (1-i) \ket{V})$ \textbf{b}. The axes of the Dirac distribution are the mutually unbiased bases $H/V$ and $D/A$; the density matrix describes the state in terms of only one basis $H/V$. Red towers indicate negative values.}
\label{fig:density}
\end{figure}

Figure~\ref{fig:density} shows directly measured Dirac distributions and the corresponding density matrices of different polarisation states. A variety of states are created for calibration as in the first experiment, but here all outcomes of the strong measurement are considered.  

\vspace{0.3cm}{\raggedleft {\it Discussion.$-$}} Figure \ref{fig:weakvals} shows that the largest divergence between theory and result in experiment 1 occurs when the initial state is anti-diagonal and therefore orthogonal to the post-selected state of diagonal. In this limit, the weak value is undefined, whereas the pointer reaches a maximum displacement (see \cite{Aharonov:1988wha,Geszti:2010ju}).
This difficulty is overcome by the full state characterisation technique performed in experiment 2. Recalling equation (\ref{Diracdistribution}), it is precisely in this regime where $p_{b_j}$ goes to zero, cancelling the effect of the breakdown of the weak-value approximation.

We note that density matrices determined by the technique demonstrated herein are not guaranteed to be precisely Hermitian due to measurement noise. For example, the density matrices shown in Figure \ref{fig:density} have small imaginary components along the diagonal, of magnitude on the order of the measurement uncertainty ($\approx 3\%$).

The similarity between equations (\ref{wavefunction}) and (\ref{Diracdistribution}) suggests a simple connection between the coefficients of the wavefunction and the entries of its Dirac distribution. In the case that the state is pure, we may combine the two equations to determine the real constant of proportionality that relates the two:
\begin{equation} \label{insight}
c_i = \frac{\nu}{p_{b_j}}S_{ij}.
\end{equation}
We see that there is a column $j$ of the Dirac distribution that is proportional to the wavefunction $c_i$. It is, in particular, the column corresponding to the choice of post-selection in equation (\ref{wavefunction}) that renders $\nu$ independent of $i$ (and hence $\nu$ can be taken to be real).

Equation (\ref{insight}) has particular relevance to our experiment for the states that have a constant of normalisation $\nu$ equal to unity and a probability of post-selection equal to one half. These states lie on the great circle of the Poincar\'{e} sphere that includes $\{ |H\rangle$, $|R\rangle$, $|V\rangle$, $|L\rangle \}$ (red points in Figure \ref{fig:sphere}, weak values and probability amplitudes in Supplementary Figure S1). Each state on this circle is from a basis that is mutually unbiased with respect to the strong measurement $\pi_{_D}$. For these states, we see that $c_i = 2 S_{ij}$ and hence the wavefunction is twice a column of the Dirac distribution. See Figure \ref{fig:density} for two examples.

The technique we demonstrate compares favourably with quantum tomography with regards to inferring the density matrix from measurement results.
Tomography via maximum likelihood estimation or least-squares fitting, which is an example of an inverse problem, becomes prohibitively difficult as the dimension of the state or number of particles in a multipartite state increases. The difficulty arises from the computational requirements of varying the vast number of fit parameters needed to estimate the state \cite{Agnew:2011wj,Agnew:2012gs}.
In contrast, no fitting is required to determine the density matrix from the directly measured Dirac distribution because it is calculated analytically.
Hence, we anticipate that for high-dimensional quantum systems especially, direct measurement will become a widely used technique for quantum state determination.

The technique we present has several logical extensions, such as directly measuring polarisation of single photons or multipartite states. Since, for the case of coherent states, the photon detection amplitude is analogous to the classical Maxwell field, the main difference between our experiment and the single-photon experiment is one of a technical nature.  One possibility is to measure the spatial distribution of the single photons with a cooled CCD \cite{Kocsis:2011jg} or electron-multiplying CCD \cite{Edgar:2012bl}.  In order to measure the Dirac distribution describing polarisation-entangled photons, our weak and strong measurement schemes would be duplicated for the signal and idler photons, together with a multiplexed coincident measurement scheme.  The required sixteen post-selection probabilities and joint weak values \cite{Lundeen:2005jr} can be established by measuring all four combinations of position and momentum of both pointers, for each of the four post-selection outcomes. This could be achieved with presently available technology using screening slits in the appropriate planes, and triggered bucket detectors.

Direct measurement can also be extended to study other discrete systems, such as the coupled spin qubits that exist in solid-state implementations of quantum information experiments \cite{Simmons:2011ik, Thewalt:2012us}.
The entire direct measurement process can be viewed as a quantum circuit, where the weak interaction is viewed as an entangling operation between the pointer and probe (see, for example, Ref. \cite{DiLorenzo:2012kv}). This means the complex-valued description of an unknown state can be determined and used within the context of a larger quantum algorithm.
Additionally, the relationship between the number of photons and the associated uncertainty of the measured state is an open question, and this is an area of current investigation.

\vspace{0.3cm}{\raggedleft {\it Conclusion.$-$} } In conclusion, we have performed the first direct measurements of general polarisation states of light.   We obtain our results through parallel measurements of the real and imaginary components of the weak value of polarisation.   An important result is that a single weak value, corresponding to the weak measurement of only one observable, determines both complex coefficients of the pure state of a qubit.  We provide some ideas for plausible extensions to this work. Direct measurement using weak values is poised to be a very promising alternative to quantum tomography. This is especially the case in discrete high-dimensional systems or experiments where the quantum state must be recorded directly by the apparatus.

\section*{Methods}
A near-Gaussian pointer state is prepared by passing HeNe laser light through a single mode fiber (SMF). The probe is then prepared by polarising the light with a polarising beam splitter (PBS) then rotating the angle of polarisation with a half-wave plate ($\lambda/2$) and/or quarter-wave plate ($\lambda/4$). 

The weak measurement is performed by coupling polarisation information to the spatial degree of freedom of the light. Light incident at an oblique angle on an $X$-cut quartz crystal undergoes a polarisation-dependent parallel displacement. By aligning the extraordinary axis with the $x$-axis and rotating the crystal about the $y$-axis, horizontal and vertical polarisations become slightly separated in $x$. We take the $z$-axis to be the direction of propagation of the light, $x$ to be the transverse direction parallel to the optical table, and $y$ to be the transverse direction perpendicular to the table. The angle of incidence was adjusted to $\sim40^{\circ}$ to ensure the two optical paths through the crystal are equal, mod $2\pi$, through the ordinary and extraordinary axes (based on crystal thickness of 700 $\mu$m). 

It is important that the pointer state be a Gaussian with a flat phase-front. We collect the SMF output with a microscope objective ($10\times$), and focus the light onto the quartz crystal ($\approx 45$ cm away). This ensures the phase-front is approximately flat over the region of interaction with the quartz.

After strong measurement, the real part of the weak value is proportional to the average position $\langle x \rangle$ of the post-selected intensity distribution immediately behind the quartz. The quartz plane is imaged onto the camera by two sets of relay optics. The first set images ($2f_1$-$2f_2$ imaging system, $f_1 = 100$ mm, $f_2 = 125$ mm) to a spatial filter (adjustable iris) that allows us to eliminate back-reflections created in the quartz crystal; the second set images ($2f_1$-$2f_2$ imaging system, $f_1 = 75$ mm, $f_2 = 250$ mm) the iris plane onto the camera. The imaginary part of the weak value is proportional to the average position of the intensity distribution in the far-field $\langle p_x \rangle$ of the quartz plane. A Fourier transform lens ($f = 300$ mm) maps the far-field distribution of the iris plane onto the camera.

We establish the expectation value of each pointer by first integrating each intensity distribution $I(x,y)$ along $y$ to find $I(x) = \sum_{y\,\mathrm{pixels}} I(x,y) \Delta y$, followed by finding the average $\langle x \rangle = \sum_{x\,\mathrm{pixels}} x I(x) \Delta x / \sum_{x\,\mathrm{pixels}} I(x) \Delta x$. This procedure is repeated with the image of the far-field to establish $\langle p_x\rangle$, and for each strong measurement outcome.

The expectation values $\langle x \rangle$ and $\langle p_x \rangle$ of the pointer, and their corresponding standard deviations, are established by averaging 100 CCD images, each with a 2000-$\mu$s exposure time. The only exception is for the data used to calibrate the weak values for Figure \ref{fig:density}b, where we averaged over 50 CCD images, each with a 500-$\mu$s exposure time. This was to reduce the effect of spot drift over the course of the calibration run where many states are measured sequentially.

A simple background subtraction is performed before calculating the pointer's position and momentum. We subtract the value of the minimum pixel from all pixels on each exposure, to reduce the effect that the background has on calculating the average. For the post-selection probability measurements used to determine the Dirac distribution, 
background subtraction is performed for each region of interest by subtracting the recorded intensity when the laser is blocked. The intensity after background subtraction of the near-field image corresponding to the outcome $|D\rangle$ is $I_D$ and of $|A\rangle$ is $I_A$. Thus the probabilities are calculated according to $p_D = I_D/(I_D + I_A)$ and $p_A = I_A/(I_D+I_A)$.

The weak value is obtained from average pixel number by
\begin{equation}
\langle \pi_{_H} \rangle^{W}_{D} = a \langle x \rangle - b + i (c \langle p_x \rangle - d),
\end{equation}
where $a,b,c,d$ are constants that must be determined by calibrating the measurement apparatus. Another set of calibration constants $a',b',c',d'$ must be determined for the post-selection of $|A\rangle$ to convert average pixel to $\langle \pi_{_H} \rangle^{W}_{A}$. We perform calibrations of the measurement apparatus by measuring the wavefunctions and Dirac distributions of known pure states and comparing $\langle x \rangle$ and $\langle p_x \rangle$ to theoretically calculated weak values.

\bibliographystyle{plainnat}

\begin{thebibliography}{10}

\bibitem{DAriano:2003tw}
D'Ariano, G.~M., Paris, M.~G.~A. \& Sacchi, M.~F. Quantum tomography.  \textit{
  Adv. Imaging Electron Phys.}, \textbf{128,} 205-308 (2003).

\bibitem{Banaszek:1999ur}
Banaszek, K., D'Ariano, G.~M., Paris, M.~G.~A. \&  Sacchi, M.~F.
   Maximum-likelihood estimation of the density matrix.  \textit{ Phys. Rev. A},
  \textbf{61,} 010304 (1999).

\bibitem{White:2001vi}
White, A.~G., James, D.~F.~V., Munro, W.~J. \&  Kwiat, P.~G.  Exploring Hilbert
  space: accurate characterization of quantum information.  \textit{ Phys.
  Rev. A}, \textbf{65,} 012301 (2001).

\bibitem{Itatani:2004wp}
Itatani, J. \textit{et al.}  Tomographic imaging of molecular
  orbitals.  \textit{ Nature}, \textbf{432,} 867-871 (2004).

\bibitem{Resch:2005vj}
Resch, K.~J., Walther, P. \& Zeilinger, A.  Full characterization of a
  three-photon Greenberger-Horne-Zeilinger state using quantum state
  tomography.  \textit{ Phys. Rev. Lett.}, \textbf{94,} 70402 (2005).

\bibitem{Agnew:2011wj}
Agnew, M., Leach, J.,McLaren, M., Roux, F.~S. \& Boyd, R.~W.  Tomography of the
  quantum state of photons entangled in high dimensions.  \textit{ Phys. Rev. A},
  \textbf{84,} 062101 (2011).

\bibitem{Lundeen:2011hj}
Lundeen, J.~S., Sutherland, B., Patel, A., Stewart, C. \& Bamber, C.  Direct
  measurement of the quantum wavefunction.  \textit{ Nature}, \textbf{474,}
  188-191 (2011).

\bibitem{Lundeen:2012db}
Lundeen J.~S. \& Bamber, C.  Procedure for direct measurement of general
  quantum states using weak measurement.  \textit{ Phys. Rev. Lett.}, \textbf{108,}
  70402 (2012).

\bibitem{Dirac:1945wx}
Dirac, P.~A.~M.  On the analogy between classical and quantum mechanics. 
  \textit{ Rev. Mod. Phys.}, \textbf{17,} 195-199 (1945).

\bibitem{Chaturvedi:2006kz}
Chaturvedi, S. \textit{et al.}
   {Wigner-Weyl correspondence in quantum mechanics for continuous and
  discrete systems---a Dirac-inspired view},  \textit{ J. Phys. A}, \textbf{39,} 1405-1423 (2006).
  
\bibitem{Hofmann:2012ey}
 Hofmann, H. F. {Complex joint probabilities as expressions of reversible transformations in quantum mechanics.} \textit{New J. Phys.}, \textbf{14,} 043031 (2012).

\bibitem{Feynman:1987ur}
Feynman, R.~P. {Negative probability.} In \textit{Quantum implications: Essays in Honour of David Bohm} by Healy, B.~J. \& Peat, F.~D. (Routledge, 1987).

\bibitem{Leonhardt:1995wb}
Leonhardt, U. {Quantum-state tomography and discrete Wigner function.} \textit{Phys. Rev. Lett.}, \textbf{74,} 4101-4105 (1995).

\bibitem{Mukamel:2003uv}
Mukamel, E., Banaszek, K., Walmsley, I. \& Dorrer, C. {Direct measurement of the spatial Wigner function with area-integrated detection.} \textit{Opt. Lett.}  \textbf{28,} 1317-1319 (2003).

\bibitem{Smith:2005vj}
Smith, B.~J., Killett, B., Raymer, M., Walmsley, I. \& Banaszek, K. {Measurement of the transverse spatial quantum state of light at the single-photon level.} \textit{Opt. Lett.}  \textbf{30,} 3365-3367 (2005).

\bibitem{Aharonov:1988wha}
Aharonov, Y., Albert, D.~Z. \& Vaidman, L.  How the result of a measurement of
  a component of the spin of a spin-1/2 particle can turn out to be 100.  \textit{
  Phys. Rev. Lett.}, \textbf{60,} 1351-1354 (1988).

\bibitem{Duck:1989vo}
Duck, I.~M., Stevenson, I.~M. \&  Sudarshan, E.~C.~G.  The sense in which a
   weak measurement  of a spin-1/2 particle's spin component yields a value
  100.  \textit{ Phys. Rev. D}, \textbf{40,} 2112-2117 (1989).

\bibitem{Knight:1990vba}
Knight, J.~M. \& Vaidman  L.  Weak measurement of photon polarization.  \textit{
  Phys. Lett. A}, \textbf{143,} 357-361 (1990).

\bibitem{Ritchie:1991vf}
Ritchie, N.~W.~M., Story,  J.~G. \& Hulet, R.~G.  Realization of a measurement
  of a  weak value. \textit{ Phys. Rev. Lett.}, \textbf{66,} 1107-1110 (1991).

\bibitem{Wiseman:2002jt}
Wiseman, H.,  {Weak values, quantum trajectories, and the cavity-QED experiment
  on wave-particle correlation.}  \textit{ Phys. Rev. A}, \textbf{65,} 032111 (2002).

\bibitem{Solli:2004ix}
Solli, D.~R., McCormick,  C.~F., Chiao, R.~Y., Popescu, S. \& Hickmann,  J.~M.
   Fast light, slow light, and phase singularities: A connection to
  generalized weak values.  \textit{ Phys. Rev. Lett.}, \textbf{92,} 043601 (2004).

\bibitem{Resch:2004es}
Resch, K.~J., Lundeen,  J.~S. \& Steinberg,  A.~M.  Experimental realization of
  the quantum box problem.  \textit{ Phys. Lett. A}, \textbf{324,} 125-131 (2004).

\bibitem{Johansen:2004cc}
Johansen, L.  {Weak Measurements with Arbitrary Probe States.}  \textit{ Phys.
  Rev. Lett.}, \textbf{93,} 120402 (2004).

\bibitem{Pryde:2005gk}
Pryde, G.~J., O'Brien, J.~L., White,  A.~G., Ralph, T.~C. \& Wiseman,  H.~M.
   Measurement of quantum weak values of photon polarization.  \textit{ Phys.
  Rev. Lett.}, \textbf{94,} 220405 (2005).

\bibitem{Hosten:2008ih}
Hosten, O. \& Kwiat, P.  Observation of the spin hall effect of light via weak
  measurements.  \textit{ Science}, \textbf{319,} 787-790 (2008).

\bibitem{Dixon:2009cz}
Dixon, P.~B., Starling, D.~J., Jordan,  A.~N. \& Howell,  J.~C.  Ultrasensitive
  beam deflection measurement via interferometric weak value amplification.
  \textit{ Phys. Rev. Lett.}, \textbf{102,} 173601 (2009).

\bibitem{Popescu:2009cz}
Popescu, S. Viewpoint: Weak measurements just got stronger. \textit{ Physics}, \textbf{2,} 32 (2009).

\bibitem{Kocsis:2011jg}
Kocsis, S. \textit{et al.}  Observing the average trajectories of single photons
  in a two-slit interferometer.  \textit{ Science}, \textbf{332,} 1170-1173 (2011).

\bibitem{Feizpour:2011bs}
Feizpour, A., Xing, X. \& Steinberg,  A.~M.  Amplifying single-photon
  nonlinearity using weak measurements.  \textit{ Phys. Rev. Lett.}, \textbf{107,}
  133603 (2011).

\bibitem{deGosson:2012kt}
de~Gosson, M.~A. \& de~Gosson,  S.~M.  {Weak values of a quantum observable and
  the cross-Wigner distribution.}  \textit{ Phys. Lett. A}, \textbf{376,} 293-296
  (2012).
 
\bibitem{Edgar:2012bl}
Edgar, M.~P. \textit{et al.} {Imaging high-dimensional spatial entanglement with a camera.} \textit{Nature Commun.}, \textbf{3,} 984 (2012).

\bibitem{Lundeen:2005jr}
Lundeen J.~S. \& Resch,  K.~J.  {Practical measurement of joint weak values
  and their connection to the annihilation operator.}  \textit{ Phys. Lett. A},
  \textbf{334.} 337-344  (2005).

\bibitem{Wootters:1989uo}
Wootters, W.~K. \& Fields,  B.~D.  Optimal state-determination by mutually
  unbiased measurements.  \textit{ Ann. Phys.}, \textbf{191,} 363-381
  (1989).
 
\bibitem{Geszti:2010ju}
Geszti, T.  Postselected weak measurement beyond the weak value.  \textit{ Phys.
  Rev. A}, \textbf{81,} 044102 (2010).

\bibitem{Agnew:2012gs}
Agnew, M., Leach, J. \& Boyd,  R.~W.  {Observation of entanglement witnesses
  for orbital angular momentum states.}  \textit{ Eur. Phys. J. D}, \textbf{66,}
  156 (2012).

\bibitem{Simmons:2011ik}
Simmons, S. \textit{et al.} {Entanglement in a solid-state spin ensemble.} \textit{Nature}, \textbf{470,} 69-72 (2011).

\bibitem{Thewalt:2012us}
Steger, M. {Quantum Information Storage for over 180 s Using Donor Spins in a $^{28}$Si ``Semiconductor Vacuum''.} \textit{Science}, \textbf{336,} 1280-1283 (2012).

\bibitem{DiLorenzo:2012kv}
Di~Lorenzo, A. {Full counting statistics of weak-value measurement.} \textit{Phys. Rev. A}, \textbf{85,} 032106 (2012).



\end{thebibliography}

\vspace{0.3cm}

\section*{Acknowledgements}
We thank K.~Pich\'e  and F. Miatto for helpful discussions regarding this work. We thank P. B. Corkum and C. Zhang for lending us the quartz crystal. This work was supported by the Canada Excellence Research Chairs (CERC) Program.  In addition, R.~W.~B. acknowledges support of from the DARPA InPho program.

\section*{Author Contributions}
J.~Z.~S. initiated the study. The experiment was designed by J.~Z.~S., A.~S.~J., E.~B.~, and J.~L. The experiment was performed by J.~Z.~S., M.~A., and A.~S.~J., and data analysis was performed by J.~Z.~S.  R.~W.~B. oversaw all aspects of the
project.    All authors contributed to the text of the manuscript.

\section*{Additional information}
Supplementary information is available in the online version of the paper. Reprints and permission information is available online at http://www.nature.com/reprints. Correspondence and requests for materials should be addressed to J.~Z.~S.

\section*{Competing financial interests}
The authors declare no competing financial interests.

\end{document}